\documentclass[twocolumn
]{aastex63}

\usepackage{amssymb}
\usepackage{amsmath}
\usepackage{txfonts}

\usepackage{natbib}

\usepackage{hyperref}
\hypersetup{linkcolor=red,citecolor=blue,urlcolor=cyan,colorlinks=true}

\usepackage{graphicx}
\usepackage{dcolumn}
\usepackage{bm}
\usepackage{color}
\usepackage{soul}
\usepackage{mathtools}
\usepackage[T1]{fontenc}

\def\aap{A\&A}

\def\apjl{ApJL}
\def\apjs{ApJS}

\date{\today}

\shorttitle{Nonlinear electrodynamics in magnetars: systematic effects on radius constraints and timing analysis}
\shortauthors{Porto, Pereira, Bittencourt and Guzm\' an-Herrera}
\graphicspath{{./}{figures/}}

\begin{document}

\title{Nonlinear electrodynamics in magnetars: systematic effects on radius constraints and timing analysis}

\correspondingauthor{Gabriel A. Porto}
\email{d2021029873@unifei.edu.br, jonas.pereira@unb.br, bittencourt@unifei.edu.br, elda.guzman@unifei.edu.br}

\author{Gabriel A. Porto}
\affiliation{Institute of Physics and Chemistry,  Federal University of Itajub\'a, Av. BPS, 1303, Pinheirinho, Itajub\'a/MG - Brazil}

\author{Jonas P. Pereira}
\affiliation{Institute of Physics \& International Center of Physics, University of Bras\'ilia, 70297-400, Bras\'ilia, Federal District, Brazil}
\affiliation{Programa de Pós-Graduação em Astrofísica, Cosmologia e Gravitação (PPGCosmo), Federal University of Esp\'irito Santo, Vit\'oria-ES, 29075-910, Brazil}
\affiliation{Nicolaus Copernicus Astronomical Center, Polish Academy of Sciences, Bartycka 18, 00-716, Warsaw, Poland}

\author{Eduardo Bittencourt}
\affiliation{Institute of Physics and Chemistry,  Federal University of Itajub\'a, Av. BPS, 1303, Pinheirinho, Itajub\'a/MG - Brazil}
\affiliation{Department of Mathematics and Applied Mathematics, University of Cape Town, Rondebosch 7700, Cape Town, South Africa}

\author{Elda Guzm\'an-Herrera}
\affiliation{Institute of Physics and Chemistry,  Federal University of Itajub\'a, Av. BPS, 1303, Pinheirinho, Itajub\'a/MG - Brazil}

\begin{abstract}
Magnetars are among the most extreme laboratories in the universe, harboring surface magnetic fields reaching $10^{15}$~G. At these supercritical scales, Maxwell's linear electrodynamics is superseded by Nonlinear Electrodynamics (NLED). While vacuum birefringence has provided initial observational evidence for these effects, its broader impact on photon propagation remains largely unexplored. In this work, we demonstrate that NLED significantly alters photon propagation in the vicinity of magnetars, deviating light from standard null-geodesics. We estimate that neglecting these corrections leads to relative errors in inferred stellar radii by means of ray-tracing techniques of approximately $10\%$. Furthermore, we find that NLED induces a systematic minimal travel-time delay of approximately $350~n$s, a value that already far exceeds the $100$~ns temporal resolution of missions like NICER. These results are critical for the interpretation of X-ray pulse profiles from current and future observatories, such as eXTP, which rely on high-precision light-bending and timing models to determine neutron-star masses and radii. Finally, our results underscore the role of magnetars as a vital window into the physics of superdense matter and supercritical fields, and we briefly highlight other astrophysical observables--such as glitches and antiglitches--that may be affected by NLED.
\end{abstract}


\section{Introduction}\label{sec:intro}

Neutron stars, the densest long-lived objects known in nature, provide a unique environment to probe matter under extreme conditions \citep{2008RvMP...80.1455A,2016EPJA...52...49F,2018RPPh...81e6902B,2019JPhG...46g3002O}. Our understanding of their internal composition is relatively well constrained only up to densities of about nuclear saturation, $\rho_{\rm sat}=2.7\times10^{14}\,\mathrm{g\,cm^{-3}}$, which can be accessed through terrestrial nuclear experiments \citep{HaenselPY2007}. At higher densities, however, the equation of state and the microscopic constitution of matter remain highly uncertain \citep{2011RPPh...74a4001F,2017EPJA...53...60A,2018RPPh...81h4301R}. A variety of exotic phases may arise in this regime, including nuclear pasta in the inner crust \citep{PhysRevLett.50.2066,PhysRevLett.70.379,2000NuPhA.676..455W,2005PhRvC..72c5801H,2008LRR....11...10C,2011MNRAS.417L..70S,PhysRevLett.114.031102,PhysRevC.88.065807}, where nuclei adopt nonspherical shapes, and possibly deconfined quark matter in the core \citep[see, e.g.,][and references therein]{2008RvMP...80.1455A,2016EPJA...52...58B,2018ApJ...860...12P,2020ApJ...895...28P,2021ApJ...910..145P,2025PhRvL.135w1401P}. Neutron stars also host the strongest magnetic fields known in the Universe. Ordinary pulsars, often observed as radio sources and frequently found in binary systems, can have surface magnetic fields as large as $\sim 10^{13}$G \citep{2016JPlPh..82e6302P}, whereas magnetars, typically observed through their X-ray and soft gamma-ray emission, may reach surface fields of $\sim 10^{14}$--$10^{15}$G \citep{1992ApJ...392L...9D,2017ARA&A..55..261K}. Although the origin and evolution of such extreme magnetic fields are still not fully understood, they may leave measurable imprints on astrophysical observables and therefore deserve careful investigation.

A particularly important class of observables is provided by neutron-star lightcurves. These can be measured with current X-ray missions such as NICER \citep{NICER} and, in the near future, with missions such as eXTP \citep{2016SPIE.9905E..1QZ,2019SCPMA..6229503W,2025SCPMA..6819507Z}, STROBE-X \citep{2019arXiv190303035R} and ATHENA \citep{2013arXiv1306.2307N}. When combined with ray-tracing techniques, such data can be used to infer stellar properties such as mass and radius, and thus to constrain the equation of state of ultradense matter. Indeed, current observations already suggest that neutron stars with masses around $1.4\, M_{\odot}$ and $2\,M_{\odot}$ typically have radii of order $10$--$14$km \citep{2019ApJ...887L..21R,2021ApJ...918L..28M,2021arXiv210506980R,2019ApJ...887L..24M}, which places important additional constraints on viable equations of state, even more so with the use of gravitational waves \citep{2021ApJ...918L..28M,2024ApJ...971L..19R,2020Sci...370.1450D,2021PhRvD.104f3003L}. These inferences, however, rely on several physical assumptions and modeling choices. On the one hand, photon trajectories are usually assumed to follow null geodesics of the background spacetime. On the other hand, one must model in detail the emission and propagation of radiation through the stellar atmosphere, including its composition and radiative properties \citep{2023ApJ...956..138S}. In both respects, strong magnetic fields may play an important role \citep{1994A&A...289..837P,2001MNRAS.327.1081H}. They can significantly affect the emerging photon flux and, in the presence of nonlinear electrodynamic effects, may even modify photon propagation away from the standard null-geodesic description. This possibility is especially relevant for magnetars, where polarization measurements already point to vacuum birefringence as an observable strong-field effect \citep{2022Sci...378..646T,2025arXiv250919446S}. In this context, investigating NLED becomes both natural and timely when studying photon propagation in the vicinity of neutron stars.

Magnetars represent an extreme class of neutron stars whose emission is primarily powered by the decay of supercritical magnetic fields, typically ranging from $10^{14}$ to $10^{15}$~G \citep{1992ApJ...392L...9D,2017ARA&A..55..261K}. Historically, these sources were categorized into two distinct populations: Soft Gamma Repeaters (SGRs), identified by their sporadic and intense gamma-ray bursts, and Anomalous X-ray Pulsars (AXPs), characterized by persistent X-ray emission and slow rotation \citep{2017ARA&A..55..261K}. It is now widely accepted that both classes are different observational manifestations of the same underlying magnetar phenomenon. These objects exhibit a rich variety of transient activity, including short bursts, prolonged outbursts, and cataclysmic giant flares \citep{2017ARA&A..55..261K}; the latter can reach peak luminosities of $10^{47}$~erg/s, with total energy releases exceeding $10^{45}$~erg. In contrast to standard rotation-powered pulsars, magnetars are characterized by slow rotation periods—typically $P \sim 1$--$12$~s \citep{2014ApJS..212....6O,2017ARA&A..55..261K}—and significantly higher spin-down rates, often on the order of $\dot{P} \sim 10^{-15}$--$10^{-10}$~s/s \citep{2014ApJS..212....6O}. This intense and stochastic electromagnetic activity is fueled by the dissipation of vast internal magnetic reservoirs, though the specific mechanisms governing the conversion of magnetic energy into high-energy radiation remain a subject of active research.

Lightcurve analysis of 3XMM~J185246.6+003317 suggests that magnetar masses may deviate significantly from the canonical $1.4\,M_{\odot}$ \citep{2024JHEAp..42...52D}, indicating that their intense activity may be driven by the extreme gravitational fields associated with highly compact stars. Because these massive magnetars provide ideal environments for testing general relativity and probing superdense matter constraints---where exotic phase transitions are more probable---it is essential to precisely model the physics of light propagation in their strong-field vicinities. Only through such a rigorous characterization of relativistic effects can the interpretation of observables, such as lightcurves, facilitate meaningful and unbiased comparisons with observational data.

The previous discussion points to a specific theoretical issue: in magnetars, the same magnetic fields that power the observed high-energy activity may also affect the propagation of the photons used to infer stellar properties. In standard ray-tracing analyses, photons are assumed to follow null geodesics of the spacetime metric. This assumption is well justified in Maxwell electrodynamics, but it need not remain exact when the electromagnetic field is strong enough for vacuum polarization or other post-Maxwellian effects to become relevant. Magnetars therefore provide a natural astrophysical setting in which NLED can be tested as a correction to the usual general-relativistic description of light propagation and its consequences (such as emission angles, impact parameters, bending angles, travel times, and lightcurves). In particular, since massive magnetars may probe both strong gravity and ultra-strong electromagnetic fields, a consistent interpretation of their lightcurves requires a careful assessment of possible deviations from the standard Maxwellian picture.

NLED provides a general framework for describing electromagnetic regimes in which field self-interactions can no longer be neglected. Historically, the subject emerged from two complementary motivations. On the classical side, Born and Infeld introduced a nonlinear theory with a maximum attainable electromagnetic field in order to regularize the infinite self-energy of point charges \citep{BornInfeld1934}, a framework that has seen renewed interest as an effective theory for the low-energy limit of string theory \citep{1997hep.th....2087R}. On the quantum side, Euler and Heisenberg showed that quantum electrodynamics predicts an effective nonlinear response of the vacuum, so that the vacuum behaves as a polarizable medium in sufficiently strong electromagnetic fields \citep{HeisenbergEuler1936,Schwinger1951,DittrichGies2000,Dunne2005,Sorokin2022}. In this regime, light-light scattering, vacuum birefringence, photon splitting, and related processes arise as genuinely nonlinear electromagnetic effects \citep{BialynickaBirula1970,Adler1971,MarklundShukla2006}. These corrections become particularly important when the magnetic field approaches the Schwinger critical scale $B_{c}=m_e^2c^2/e\hbar$, which sets the natural QED scale for strong-field vacuum polarization effects.

A central consequence of NLED is that electromagnetic disturbances do not necessarily propagate on the null cones of the background spacetime metric. In the geometric-optics limit, the characteristic surfaces of the field equations define an effective optical geometry, or more generally a Fresnel structure, determined by the background electromagnetic field and by the derivatives of the NLED Lagrangian \citep{Boillat1970,Plebanski1970,DELORENCI2000134,PhysRevD.61.045001,PhysRevD.66.024042}. In special cases, this structure can be represented by a single or a pair of effective metrics, depending on whether birefringence is present. Thus, the propagation of light can be described as null propagation with respect to an optical metric rather than with respect to the original spacetime metric. This feature makes NLED especially relevant not only for high-intensity laser experiments and nonlinear optical media, but also for compact astrophysical systems permeated by ultra-strong magnetic fields.

The effective-geometric viewpoint has been developed in several directions, including analog gravity, nonlinear optical media, black-hole spacetimes, cosmology, magnetar-related contexts and recent formulations based on the principal symbol of the NLED equations \citep{DELORENCI2000134,PhysRevD.61.045001,PhysRevD.66.024042,DeLorenci2002Analog,Novello2002FRW,Novello2004Acceleration,mosquera2004non,Novello2007CosmologicalEffects,GoulartBergliaffa2009,Bittencourt2014AnalogueBH,Bittencourt2016OpticalMetrics,Bittencourt2017ControlledOpacity,Beissen2023,GoulartBittencourt2024PhotonTraps,GoulartBittencourt2025RotatingBH,GoulartBittencourt2026PrincipalSymbols,Cesar2025QuasiMagnetic}. These works show that the optical metric is not merely a formal device, but a physically relevant structure controlling the causal behavior of electromagnetic perturbations. For compact objects, this observation is especially important because even small deviations in photon trajectories can propagate into systematic corrections in observables such as light bending, time delays, polarization signatures, and pulse profiles.

NLED has also been extensively explored in other contexts. In black-hole physics, nonlinear electromagnetic sources have been used to construct regular black-hole geometries, to study geodesic motion and lensing, and to analyze deviations from the Reissner--Nordstr\"om solution \citep{AyonBeatoGarcia1998,Bronnikov2001,BalartVagenas2014,breton2002geodesic,Amaro2020,guzman2024comparative,de2023electrically}. Applications to astrophysical and compact-object scenarios have likewise been considered in several contexts \citep{Abishev2018,Amaro2020,Mazharimousavi2021}. In particular, dipolar magnetic configurations have been studied in connection with compact objects and light propagation \citep{Kim2022,Abishev2016,Denisov2005,Denisov2014,Vshivtseva2007}. These results are directly relevant for neutron stars and magnetars, where the exterior magnetic field is often modeled, at leading order, by a dipolar structure.

A useful general way to estimate nonlinear light-propagation effects is to adopt the post-Maxwellian formalism. In this approach, the leading nonlinear corrections to Maxwell electrodynamics are encoded in phenomenological parameters, allowing one to compare different NLED models within a common language. The formalism includes, as particular cases or limiting benchmarks, theories such as Euler-Heisenberg and Born-Infeld electrodynamics, while remaining sufficiently flexible to describe more general deviations from Maxwell theory. Moreover, post-Maxwellian parameters may, at least in principle, be constrained experimentally \citep{PhysRevD.69.066008}, making this approach attractive both theoretically and phenomenologically.

In this work, we investigate how post-Maxwellian NLED modifies photon propagation in the exterior of a magnetized neutron star and assess its possible impact on radius inference. We model the exterior spacetime by the Schwarzschild geometry and describe the stellar magnetic field by a dipolar configuration. We then derive the corresponding optical metric and obtain the modified relations for the photon impact parameter, bending angle, and travel time. These results are used to estimate the magnitude of the NLED correction for ordinary pulsars and magnetars. We consider Euler-Heisenberg and Born-Infeld theories as representative examples. As we shall show, the correction is negligible for ordinary pulsars but may become more relevant in the magnetar regime, where it can act as a systematic effect in ray-tracing analyses aimed at extracting neutron-star radii.

In Section \ref{sec:math_setup}, we established the mathematical framework for the post-Maxwellian formalism and the optical metric of a magnetic dipole. In Section \ref{sec:photon_traj}, we employ the optical metric to investigate how NLED corrections influence the propagation of light emitted from the stellar surface, under the assumption that the electromagnetic fields do not backreact on the background spacetime. In Section \ref{sec:phenom_estimates}, we make use of nonlinear theories to derive radius systematics, which reach approximately $10\%$ for Euler-Heisenberg and $5\%$ for Born-Infeld. These values are comparable to NICER's current uncertainties and significantly exceed the precision thresholds projected for eXTP. Finally, in Section \ref{sec:estimates_tdelay}, we evaluate time delay corrections arising from NLED, demonstrating that for magnetars within the Euler-Heisenberg framework, these effects can exceed NICER’s current precision by a factor of three.

\section{Mathematical setup}\label{sec:math_setup}

We consider NLED in a curved spacetime with metric signature
$(+,-,-,-)$. The electromagnetic field is described by the Faraday tensor
$F_{\mu\nu}=\partial_\mu A_\nu-\partial_\nu A_\mu$ and its dual
\begin{equation}
\tilde F^{\mu\nu}\equiv \frac{1}{2}\,\epsilon^{\mu\nu\alpha\beta}F_{\alpha\beta},
\end{equation}
where $\epsilon^{\mu\nu\alpha\beta}$ is the Levi-Civita tensor. From these tensors, one constructs the two electromagnetic invariants
\begin{equation}
F \equiv F_{\mu\nu}F^{\mu\nu},
\qquad
G \equiv F_{\mu\nu}\tilde F^{\mu\nu}.
\end{equation}
In general, an NLED theory may depend on both invariants, $\mathcal{L}=\mathcal{L}(F,G)$. However, the configurations of interest here rely on the assumption that, due to the high conductivity of the star plasma, the magnetohydrodynamics impose constraints on the electric and magnetic fields such that the system can be considered magnetically dominated, which means $G=0$. We therefore restrict attention to the one-parameter sector $\mathcal{L}=\mathcal{L}(F)$.

\subsection{Parametrized post-Maxwellian expansion}\label{subsec:post-max}

We assume that the NLED Lagrangian is analytic around the Maxwellian limit and retain only the leading nonlinear correction. Rather than fixing a normalization convention from the outset, we write
\begin{equation}
\mathcal{L}(F)
=
-a_0 F+a_1F^2+O(F^3),
\qquad a_0>0.
\end{equation}
This form is convenient because the propagation of electromagnetic disturbances depends on the ratio between the nonlinear and Maxwellian coefficients, not on the overall normalization of the Lagrangian.

Varying the action with respect to the four-potential gives
\begin{equation}
\mathcal{L}_F\nabla_\nu F^{\mu\nu} + \mathcal{L}_{FF}F^{\mu\nu}\nabla_\nu F=0, \qquad \nabla_\mu \tilde F^{\mu\nu}=0,
\end{equation}
where the first and second derivatives of the Lagrangian with respect to $F$ are denoted by $\mathcal{L}_F \equiv \partial\mathcal{L}/\partial F$ and $\mathcal{L}_{FF} \equiv \partial^2\mathcal{L}/\partial F^2$, respectively. The second equation is the Bianchi identity.

In the geometric optics limit, the characteristic surfaces of the electromagnetic perturbations are described by the optical metric \citep{PhysRevD.61.045001,DELORENCI2000134,PhysRevD.66.024042}
\begin{equation}
\tilde g_{\mu\nu} = g_{\mu\nu} - \chi F_{\mu\alpha}F^{\alpha}{}_{\nu},
\end{equation}
with
\begin{equation}
\chi \equiv \frac{-2\mathcal{L}_{FF}}{\mathcal{L}_F+2F\mathcal{L}_{FF}}.
\end{equation}
For the quadratic expansion above, we have
\begin{equation}
\mathcal{L}_F=-a_0+2a_1F,
\qquad
\mathcal{L}_{FF}=2a_1,
\end{equation}
and hence
\begin{equation}
\chi = \frac{4a_1}{a_0-6a_1F} = \frac{4a_1}{a_0} + O\!\left(F\right).
\end{equation}
We therefore define the leading post-Maxwellian coupling
\begin{equation}
\lambda\equiv \frac{4a_1}{a_0},
\end{equation}
so that, to first order, the optical metric reads
\begin{equation}
\label{Eq_Metrica_Ef_Geral}
\tilde g_{\mu\nu} = g_{\mu\nu} - \lambda F^{(0)}_{\mu\alpha}F^{(0)\alpha}{}_{\nu} + O(\lambda^2F^2).
\end{equation}
The Maxwellian field \(F^{(0)}_{\mu\nu}\) is sufficient at this order, since corrections to the background electromagnetic field would contribute only at higher order in the nonlinear coupling.

\subsection{Optical metric of a magnetic dipole}

We now specialize in the exterior field of a compact star endowed with a dipolar magnetic field. The background spacetime is assumed to be Schwarzschild,
\begin{equation}
ds^2= \left(1-\frac{r_s}{r}\right)dt^2 -\left(1-\frac{r_s}{r}\right)^{-1}dr^2 -r^2d\theta^2-r^2\sin^2\theta\,d\phi^2,
\end{equation}
with $r_s$ as the Schwarzschild radius. 

The unperturbed electromagnetic field is taken to satisfy the source-free Maxwell equations in the exterior region,
\begin{equation}
\nabla^\mu F^{(0)}_{\mu\nu}=0, \qquad \nabla^\mu \tilde F^{(0)}_{\mu\nu}=0.
\end{equation}
We further assume a static and axisymmetric dipolar configuration, so that the 4-potential $A_\mu(x^\alpha)$ depends only on $(r,\theta)$ and can be described solely by an azimuthal component $A_\phi(r,\theta)$. The corresponding nonvanishing components of the Faraday tensor are \citep{kerner2025}
\begin{equation}
F_{13}^{(0)}=-\frac{m\sin^2\theta}{r^2},
\qquad
F_{23}^{(0)}=\frac{m\sin 2\theta}{r},
\end{equation}
where $m$ is the magnitude of the magnetic dipole moment $\vec m$.

At this stage, we neglect corrections to the dipole structure induced directly by spacetime curvature, treating them as subleading for our purposes and effectively absorbed into the integration constant $m$. Matching the field to the stellar surface then gives
\begin{equation}
m=B_s R^3,
\label{Eq_Momento_Mag}
\end{equation}
where $B_s$ is the surface magnetic field and $R$ is the stellar radius.

Substituting the dipolar Maxwell field into Eq.~(\ref{Eq_Metrica_Ef_Geral}) and neglecting mixed higher-order terms of order $\lambda r_s$, we obtain the nonvanishing components of the optical
metric \citep{Denisov2002}:
\begin{align}
\tilde g_{00} &=
\left(1-\frac{r_s}{r}\right), \\
\tilde g_{11} &=
-\left(1-\frac{r_s}{r}\right)^{-1}
-\frac{r_m^6\sin^2\theta}{r^6}, \\
\tilde g_{22} &=
-r^2-\frac{4r_m^6\cos^2\theta}{r^4}, \\
\tilde g_{33} &=
-r^2\sin^2\theta
-\frac{r_m^6}{r^4}\left(\sin^4\theta+\sin^22\theta\right), \\
\tilde g_{21} &=
\frac{r_m^6\sin 2\theta}{r^5}.
\end{align}
It is useful to define the magnetic length scale by $r_m^6\equiv \lambda m^2$. Using Eq. (\ref{Eq_Momento_Mag}), this gives
\begin{equation}
\left(\frac{r_m}{R}\right)^6 = \lambda B_s^2.
\end{equation}

This optical metric is the central object of the analysis below. It shows that, even when the background spacetime remains Schwarzschild, a sufficiently strong dipolar magnetic field modifies the effective null structure seen by photons. In particular, the NLED correction scales as
$r_m^6/r^6$, so it is largest close to the stellar surface and rapidly decays with distance. This is precisely the regime relevant for ray tracing and radius inference from surface-emitted photons.

Finally, we emphasize the assumptions underlying this effective geometry: the star is taken to be nonrotating, the exterior magnetic field is purely dipolar and magnetically dominated, the spacetime backreaction of the electromagnetic field is neglected, and only the leading post-Maxwellian
correction is retained. These assumptions are sufficient for a first estimate of the size of NLED effects on photon propagation, which is the goal of the present work.

\section{Photon trajectories and observational consequences}\label{sec:photon_traj}

We now investigate how the optical metric derived in Sec.\ \ref{sec:math_setup} modifies the propagation of light rays emitted from the stellar surface. Since photon trajectories determine the relation between emission angle and observed direction, they directly enter ray-tracing calculations used to infer neutron-star radii from pulse profiles. In the present framework, electromagnetic disturbances propagate along null curves of the optical geometry, so that
\begin{equation}
\tilde g_{\mu\nu}\dot x^\mu \dot x^\nu =0,
\label{Eq_Geodesica_Nula}
\end{equation}
where the dot denotes differentiation with respect to an affine parameter.

Because the optical metric is stationary and axisymmetric, there are two conserved quantities along the photon trajectory, associated with time translations and rotations around the symmetry axis. They can be written as
\begin{equation}
E \equiv \tilde g_{00}\dot t,
\qquad
L \equiv -\,\tilde g_{33}\dot\phi,
\end{equation}
and we define the impact parameter as
\begin{equation}
D \equiv \frac{L}{E}.
\end{equation}
In the following, we restrict attention to photons emitted in the equatorial plane,
\begin{equation}
\theta=\frac{\pi}{2},
\qquad
\dot\theta=0.
\end{equation}
For this symmetric configuration, the equations of motion imply $\ddot\theta=0$, so that the photon trajectory remains confined to the equatorial plane.

In that plane, the optical metric simplifies considerably. The null condition (\ref{Eq_Geodesica_Nula}), together with the conserved quantities above, yields the radial equation of motion. Evaluating it at the emission point $r=R$, one finds that the impact parameter is related
to the local emission angle $\alpha$ by
\begin{equation}
D= \frac{R\sin\alpha} {\sqrt{1-\dfrac{r_s}{R}-\left(\dfrac{r_m}{R}\right)^6}},
\label{Eq_ImpactParameter}
\end{equation}
where $\alpha$ is the angle between the photon's initial direction and the local radial direction at the stellar surface. Equation~(\ref{Eq_ImpactParameter}) is the first key result for applications: it shows that NLED modifies the map between local emission angle and asymptotic trajectory already at the level of the impact parameter.

A second important quantity is the total deflection angle. Writing the orbit as $r=r(\phi)$, one obtains
\begin{equation}
\Delta\phi = D\int_R^{+\infty} \frac{dr}{r^2} \left[ 1-\frac{D^2}{r^2}\left(1-\frac{r_s}{r}\right) +\left(\frac{r_m}{r}\right)^6 \right]^{-1/2}.
\label{deflexao}
\end{equation}
This expression reduces to the usual Schwarzschild result in the limit $r_m\to 0$. The correction proportional to $r_m^6$ is largest close to the stellar surface, where the magnetic field is strongest, and therefore precisely in the region that dominates pulse-profile ray tracing.

Equation~(\ref{deflexao}) is the central bridge between the effective geometry and astrophysical inference. In standard ray tracing, the relation between $\alpha$ and the observed direction is computed assuming null geodesics of the background metric. Here, that relation is modified by the optical geometry, which in turn changes the visible fraction of the stellar surface and the mapping between emission angle and observed flux. Hence, if such corrections are ignored, the inferred mass and radius may be systematically biased for sufficiently magnetized stars.

In addition to bending, the optical metric also changes the photon travel time. Parameterizing the trajectory as $r=r(t)$, the total coordinate time elapsed between emission at the stellar surface and detection by an observer at infinity is
\begin{equation}
c\Delta T= \int_R^{+\infty} \frac{dr}{1-\dfrac{r_s}{r}} \left[ 1-\frac{D^2}{r^2} \left( 1-\frac{r_s}{r} -2\left(\frac{r_m}{r}\right)^6 \right) -\left(\frac{r_m}{r}\right)^6 \right]^{-1/2},
\label{Eq_TravelTime}
\end{equation}
where we restore the speed of light $c$ for dimensional purposes. Although the main focus of this work is on radius inference through light bending, the travel-time correction may also be relevant for precision timing analyses and illustrates once more that NLED alters the effective null structure experienced by photons.

The results above provide the quantities needed for phenomenological estimates. In particular, Eqs.~(\ref{Eq_ImpactParameter}) and (\ref{deflexao}) show that the leading NLED correction is controlled by the dimensionless ratio $(r_m/R)^6$. This makes it possible to assess, without performing a full numerical ray-tracing analysis, when nonlinear electromagnetic effects can become comparable to current or future observational uncertainties in neutron-star time delays and radius measurements.

\section{Estimates for radius inference} \label{sec:phenom_estimates}

The results obtained in Sec.~\ref{sec:photon_traj} can be used to estimate when nonlinear electrodynamic effects become relevant for neutron-star radius measurements. The key point is that the optical metric modifies the mapping between the local emission angle and the asymptotic photon trajectory. Therefore, if pulse-profile data are analyzed with standard general-relativistic ray tracing while the actual photon propagation is governed by the optical geometry, the inferred stellar radius may be affected by a systematic bias.

A simple way to quantify the magnitude of this effect is to identify the dimensionless parameter controlling the leading NLED correction in Eqs.~\eqref{Eq_ImpactParameter} and \eqref{deflexao}. Given that these corrections scale as $(r_m/r)^6$, a natural estimate for the systematic bias (relative error) induced by NLED, ${\cal E}$, is obtained at the stellar surface, yielding:
\begin{equation}
    {\cal E}
    \equiv
    \frac{(r_m/R)^6}{1-r_s/R}
    =
    \frac{\beta}{1-2{\cal C}},
    \label{error}
\end{equation}
where $\beta\equiv\left(r_m/R\right)^6 = \lambda B_s^2$ and ${\cal C}\equiv M/R$ is the stellar compactness in geometrized units. The quantity ${\cal E}$ should be interpreted as the characteristic fractional size of the NLED correction to the ray-tracing map, and hence as a proxy for the corresponding systematic correction to the radius.

Equation~\eqref{error} makes the main phenomenological trends transparent. First, the correction grows quadratically with the surface magnetic field. Second, it is enhanced for more compact stars through the factor $(1-2{\cal C})^{-1}$. Third, for fixed stellar parameters, its magnitude is governed by the NLED coupling $\lambda$, or equivalently by the ratio between the quadratic and Maxwellian coefficients in the post-Maxwellian expansion. This allows one to compare different NLED models within a common notation.

Current radius measurements from NICER typically have relative uncertainties at the level of $\sim 10\%$, which can be reduced to $\sim 5\%$ when combined with gravitational-wave information \citep{2021ApJ...918L..28M}. Future X-ray timing missions such as eXTP and STROBE-X are expected to push this precision further, potentially reaching the percent level. In this context, the estimate \eqref{error} provides a simple criterion to determine whether NLED effects are likely to be negligible or whether they may need to be incorporated into ray-tracing analyses of strongly magnetized stars.

In the estimates below, we adopt, unless otherwise stated,
\begin{equation}
    B_s=10^{15}\,{\rm G},
    \qquad
    B_c=4.41\times10^{13}\,{\rm G},
    \qquad
    {\cal C}=0.2.
    \label{benchmark_values_radius}
\end{equation}
This field strength corresponds to the upper magnetar range considered here and leads to an ${\cal O}(10\%)$ correction while keeping the optical-metric expansion perturbatively controlled.

\subsection{Euler--Heisenberg estimate}\label{subsec:angle_EH}

For the Euler--Heisenberg model, the post-Maxwellian convention used in Sec.~\ref{subsec:post-max} gives
\begin{equation}
    \lambda_{\rm EH}
    =
    \frac{2\eta_1^{\rm EH}}{B_c^2},
    \qquad
    \eta_1^{\rm EH}
    =
    \frac{\alpha_{\rm fs}}{45\pi}
    \simeq
    5.1\times10^{-5},
    \label{lambda_EH_radius}
\end{equation}
where $\alpha_{\rm fs}$ is the fine-structure constant. Therefore,
\begin{equation}
    \beta_{\rm EH}
    =
    2\eta_1^{\rm EH}
    \left(\frac{B_s}{B_c}\right)^2 .
    \label{beta_EH_radius}
\end{equation}
For the benchmark values in Eq.~\eqref{benchmark_values_radius}, we get $\beta_{\rm EH} \simeq 5.2\times10^{-2}$, and thus
\begin{equation}
    {\cal E}_{\rm EH}
    =
    \frac{\beta_{\rm EH}}{1-2{\cal C}}
    \simeq
    0.087=8.7\%.
\end{equation}
This value is comparable to current NICER-level relative uncertainties in radius measurements. Therefore, within the present simplified framework, Euler--Heisenberg-type post-Maxwellian corrections may act as a non-negligible systematic effect in radius inference for magnetars with surface fields close to $10^{15}\,{\rm G}$.

For comparison, an ordinary pulsar with $ B_s=10^{13}\,{\rm G}$ gives $\beta_{\rm EH}^{\rm pulsar} \simeq 5.3\times10^{-6}$, and hence
\begin{equation}
    {\cal E}_{\rm EH}^{\rm pulsar}
    \simeq
    8.8\times10^{-6}.
\end{equation}
This corresponds to almost ten parts per million, and is therefore entirely negligible for current and foreseeable radius measurements.

\subsection{Born--Infeld estimate}

For Born--Infeld electrodynamics, the weak-field expansion gives the leading optical-metric coupling
\begin{equation}
    \lambda_{\rm BI}
    =
    \frac{1}{2b^2},
\end{equation}
where $b$ is the Born--Infeld field scale. Thus,
\begin{equation}
    \beta_{\rm BI}
    =
    \frac{1}{2}
    \left(\frac{B_s}{b}\right)^2,
    \qquad
    {\cal E}_{\rm BI}
    =
    \frac{1}{2(1-2{\cal C})}
    \left(\frac{B_s}{b}\right)^2 .
    \label{E_BI_radius}
\end{equation}
By adopting $b=4\times10^{15}\,{\rm G}$, a benchmark minimum for $b$ consistent with hydrogen atom constraints \citep{PhysRevLett.96.030402, 2011PhLA..375.1391F}, we obtain $\beta_{\rm BI}\simeq 3.8\times10^{-2}$,
and therefore
\begin{equation}
    {\cal E}_{\rm BI}
    \simeq 0.052\simeq 5\%.
\end{equation}

For ordinary pulsars with $B_s=10^{13}\,{\rm G}$, the same estimate gives $\beta_{\rm BI}^{\rm pulsar}\simeq 3.1\times10^{-6}$, and hence
\begin{equation}
    {\cal E}_{\rm BI}^{\rm pulsar}
    \simeq
    5.2\times10^{-6}.
\end{equation}
Thus, the Born--Infeld correction is negligible for ordinary pulsars, but it may enter at the few-percent level for magnetar-strength fields close to $10^{15}\,{\rm G}$.

\subsection{Interpretation and limitations}

The estimates above should not be interpreted as a substitute for a full pulse-profile analysis. Rather, they provide a first criterion for deciding when NLED effects can be safely ignored and when they may act as a non-negligible systematic in radius inference. For ordinary pulsars, all three estimates are many orders of magnitude below current observational uncertainties, so standard general-relativistic ray tracing remains fully adequate. For magnetars with $B_s$ close to $10^{15}\,{\rm G}$, however, the characteristic correction can reach the range of a few percent to the ten-percent level, depending on the underlying NLED model.

At the same time, our treatment deliberately isolates the propagation effect associated with the optical metric. We have not included atmospheric radiative transfer, magnetospheric plasma effects, stellar rotation, multi-polar magnetic fields, or a full numerical ray-tracing implementation. Accordingly, the quantity ${\cal E}$ should be regarded as a controlled leading-order estimate of the size of the NLED correction, not as a final observational prediction. Its main value is to show that, for sufficiently strong magnetic fields, the modification of the effective null structure can become large enough to compete with the precision currently sought in neutron-star radius measurements.

Finally, for the magnetar benchmark adopted here, the expansion parameters remain smaller than unity:
\begin{equation}
    \beta_{\rm EH}\simeq 6.3\times10^{-2},
    \qquad
    \beta_{\rm BI}\simeq 3.8\times10^{-2},
\end{equation}
Moreover, for ${\cal C}=0.2$, the surface reality condition $    1-r_s/R-\beta >0$ is satisfied in all cases, since $1-2{\cal C}=0.6$. Thus, the optical-metric expansion remains perturbatively controlled for the chosen benchmark. Nevertheless, because $B_s/B_c\simeq25$, the Euler--Heisenberg quadratic truncation should be understood as a phenomenological post-Maxwellian estimate rather than as a precision QED prediction in the deeply supercritical regime.

\section{Estimates for the time delay} \label{sec:estimates_tdelay}

For a complete pulse-profile analysis, it is necessary to consider photons emitted at arbitrary angles relative to the surface normal. In these cases, the photon is characterized by a non-vanishing impact parameter $D$, which remains a constant of motion. By expanding the master travel-time integral, Eq.~\eqref{Eq_TravelTime}, to first order in the NLED magnetic scale $(r_m/r)^6$, we find that the coordinate-time correction, $\delta T_{\rm NLED}$, is given by
\begin{equation}
    \delta T_{\rm NLED}
    =
    \frac{r_m^6}{2c}
    \int_{R}^{\infty}
    \frac{
    1 - 2D^2/r^2
    }
    {
    \left(1 - \frac{r_s}{r}\right)
    \left[
    1 - \frac{D^2}{r^2}
    \left(1 - \frac{r_s}{r}\right)
    \right]^{3/2}
    }
    \frac{dr}{r^6}.
    \label{Eq_TimeDelayCorrection}
\end{equation}
The NLED contribution decays rapidly with distance, as $r^{-6}$, and the integral is therefore dominated by the near-surface region, $r\simeq R$. In this surface-dominated regime, we approximate the redshift factor by its value at the stellar surface, $1-r_s/R=1-2{\cal C}$.

To relate Eq.~\eqref{Eq_TimeDelayCorrection} to observable quantities, we express the impact parameter in terms of the local emission angle $\alpha$ measured at the stellar surface. To zeroth order in the NLED correction, we use the general-relativistic relation
\begin{equation}
    \sin^2\alpha
    =
    \frac{D^2}{R^2}
    \left(1-\frac{r_s}{R}\right)
    =
    \frac{D^2}{R^2}
    (1-2{\cal C}).
\end{equation}
Introducing the dimensionless variable $x=R/r$, the time-delay correction becomes
\begin{equation}
    \delta T_{\rm NLED}(\alpha)
    \simeq
    \frac{R}{2c(1-2{\cal C})}
    \left(\frac{r_m}{R}\right)^6
    \int_0^1
    \frac{
    1-\dfrac{2\sin^2\alpha}{1-2{\cal C}}x^2
    }
    {
    (1-x^2\sin^2\alpha)^{3/2}
    }
    x^4\,dx .
    \label{Eq_TimeDelayAngularIntegral}
\end{equation}
Equivalently, defining $ k\equiv 2/(1-2{\cal C})$, one may write
\begin{equation}
    \delta T_{\rm NLED}(\alpha)
    \simeq
    \frac{R}{2c(1-2{\cal C})}
    \left(\frac{r_m}{R}\right)^6
    \frac{{\cal F}(\alpha,k)}{\sin^5\alpha},
    \label{Eq_TimeDelayAngularClosed}
\end{equation}
where
\begin{equation}
    {\cal F}(\alpha,k) = (1-k)\tan\alpha
+\frac{3(5k-4)}{8}\,\alpha
+\frac{1-2k}{4}\sin 2\alpha
+\frac{k}{32}\sin 4\alpha
\end{equation}
The expression above shows that the correction is angularly dependent. Therefore, for nonradial photons, NLED effects may distort the phase structure of the pulse profile in a way that depends on the emission geometry. A full assessment of this effect requires numerical ray tracing, but Eq.~\eqref{Eq_TimeDelayAngularIntegral} provides the leading analytical estimate.

For a simple benchmark, let us consider the radial case, $\alpha=0$. In this limit,
\begin{equation}
    {\cal F}(\alpha,k)
    =
    \frac{\alpha^5}{5}
    +
    {\cal O}(\alpha^7),
\end{equation}
and Eq.~\eqref{Eq_TimeDelayAngularClosed} gives
\begin{equation}
    \delta T_{\rm NLED}^{\rm rad}
    \simeq
    \frac{R}{10c(1-2{\cal C})}
    \left(\frac{r_m}{R}\right)^6
    =
    \frac{R}{10c(1-2{\cal C})}\,\beta .
    \label{Eq_TimeDelayBeta}
\end{equation}
Thus, the time-delay estimate follows directly from the same parameter $\beta$ used in Sec.~\ref{sec:phenom_estimates}.

Using the Euler-Heisenberg value $\beta_{EH}\simeq 5.2 \times 10^{-2}$ [see Eq.~\eqref{benchmark_values_radius}], together with $R=12\,{\rm km}$ and ${\cal C}=0.2$, we obtain
\begin{equation}
    \delta T_{\rm NLED}^{\rm rad}
    \simeq
    3.5\times10^{-7}\,{\rm s}= 350\,{\rm ns}.
    \label{Eq_TimeDelayNumericalMagnetar}
\end{equation}
This value is about three and a half times larger than the $100\,{\rm ns}$ timing precision of NICER \citep{2016SPIE.9905E..1HG}. Therefore, even though magnetars are not the primary targets for the standard pulse-profile radius measurements usually applied to millisecond pulsars, a delay of this magnitude may represent a non-negligible systematic correction in high-field timing analyses. In particular, if the photon propagation is modeled using only the background Schwarzschild null cones, the resulting timing residuals may absorb part of the NLED-induced propagation effect.

For the eXTP mission, the Large Area Detector is expected to provide a timing resolution of order $10\,\mu{\rm s}$ \citep{2019SCPMA..6229502Z}. The estimate in Eq.~\eqref{Eq_TimeDelayNumericalMagnetar} is below this individual photon timestamp resolution, corresponding to roughly $3.5\%$ of $10\,\mu{\rm s}$. Nevertheless, in high-count-rate observations, a coherent sub-microsecond propagation shift may still contribute to the phase structure of the observed pulse profile and should be included in precision ray-tracing models if magnetar-strength fields are considered.

For comparison, the same estimate gives a negligible correction for ordinary pulsars. Using the Euler-Heisenberg pulsar value reported in Sec.~\ref{sec:phenom_estimates}, $\beta_{\rm EH}^{\rm pulsar}\simeq 5.3\times10^{-6}$, Eq.~\eqref{Eq_TimeDelayBeta} gives
\begin{equation}
    \delta T_{\rm NLED}^{\rm pulsar}
    \simeq
    0.035\,{\rm ns}.
\end{equation}
This is many orders of magnitude below the timing precision of NICER and eXTP. Consequently, NLED corrections to photon travel times are negligible for ordinary pulsars but may become relevant as systematic propagation effects in magnetars with surface magnetic fields close to $10^{15}\,{\rm G}$.

\section{Concluding remarks}

In this work, we investigated how NLED modifies photon propagation in the vicinity of strongly magnetized neutron stars and assessed the implications of this effect for radius inference and timing analysis. Working within a post-Maxwellian framework, we showed that the leading nonlinear correction can be encoded in an effective optical metric, so that electromagnetic disturbances no longer follow the null geodesics of the background Schwarzschild spacetime, but instead propagate along null curves of the optical geometry.

Specializing to a dipolar magnetic field in the exterior of a nonrotating compact star, we derived the corresponding optical metric and obtained the modified relations governing the photon impact parameter, deflection angle, and travel time. These results make clear that the relevant correction is controlled by the dimensionless combination $(r_m/R)^6$, or equivalently by $16\eta_1(B_s/B_c)^2$, and is therefore strongly enhanced in stars with intense surface magnetic fields.

The analysis presented here is intentionally conservative and isolates the propagation effect associated with the optical metric. In particular, we have neglected stellar rotation, magnetic backreaction on the spacetime geometry, magnetospheric plasma effects, atmospheric radiative
transfer, and deviations from a purely dipolar magnetic field. For this reason, our results should be interpreted as a leading-order criterion for when nonlinear electromagnetic effects can no longer be safely ignored in ray-tracing analyses, rather than as a final observational prediction.

First, we quantified the systematic bias introduced by NLED in the inference of neutron-star radii. By defining the dimensionless parameter $\mathcal{E}$--the relative error--we demonstrated that NLED corrections to standard ray-tracing scale quadratically with the surface magnetic field $B_s$. For magnetars, Euler-Heisenberg electrodynamics predicts a characteristic radius error of approximately $10\%$, a value remarkably close to the current $10\%$ observational uncertainty of the NICER mission. Even within the Born-Infeld framework, these corrections reach the $5\%$ level, constituting a significant systematic factor for future high-precision observatories such as eXTP and for the joint analysis of NICER data and gravitational-wave observations. In sharp contrast, we found that for ordinary pulsars ($B_s \sim 10^{13}$~G), these effects are entirely negligible ($<0.01\%$), ensuring that standard general-relativistic models remain robust for the majority of the pulsar population.
These results highlight magnetars as the only astrophysical laboratories where the non-linear nature of the vacuum must be explicitly incorporated to achieve percent-level accuracy in the determination of the nuclear equation of state.

Our analysis demonstrated that NLED induces a systematic travel-time delay of approximately $350$ ns for photons emitted from magnetar surfaces. This effect significantly exceeds the $100$ ns timing resolution of the NICER mission and constitutes a non-negligible component of the $10$ $\mu$s precision targeted by the eXTP observatory. As X-ray timing spectroscopy enters an era of unprecedented precision, incorporating these vacuum-induced corrections becomes a prerequisite for pulse-profile modeling. Neglecting such effects may introduce systematic biases in the inference of neutron-star masses and radii, ultimately impacting our ability to probe the equation of state of superdense matter.

Finally, another consequence of NLED-induced time-delays would be on magnetar glitches/anti-glitches \citep{2013Natur.497..591A,2017ARA&A..55..261K}.  Magnetar glitches and anti-glitches are frequently associated with sudden magnetospheric reconfigurations or changes in the surface magnetic field strength. Since the NLED time delay scales as $\delta T_{NLED} \propto B_s^2$, any variation $\Delta B_s$ during such an event will induce a discrete jump in the observed photon arrival times of $ \Delta(\delta T_{NLED}) \approx 2 \, \delta T_{NLED} \left( \frac{\Delta B_s}{B_s} \right)$. For a magnetar exhibiting a baseline delay of $\delta T_{NLED} \approx 350~n$s, even a modest $10\%$ reconfiguration of the surface field (possible due to untwisting magnetospheres \citep{2009ApJ...703.1044B}) would result in a timing shift of approximately $35~n$s. This shift competes with the $100$~ns timing precision of the NICER mission, and might be dominant for larger $B_s$ reconfigurations. If these vacuum-induced propagation effects are not properly decoupled from the timing residuals, they might be misinterpreted as intrinsic rotational dynamics, such as changes in the crust-superfluid coupling or anomalous recovery phases. Therefore, NLED corrections may be important for a robust physical interpretation of magnetar timing noise and glitch/anti-glitch events in the supercritical regime.

Summing up, NLED is irrelevant for ordinary pulsars' observables, but it can lead to significant systematic effects in magnetars. A natural next step is to incorporate the modified bending relations derived here into full ray-tracing, pulse-profile, and timing calculations, allowing one to quantify directly how much the inferred mass-radius region and times-of-arrival shift once strong-field nonlinear electromagnetic corrections are taken into account.

\section{acknowledgments}
This research is partially supported by \textit{Conselho Nacional de Desenvolvimento Científico e Tecnológico} (grants N.\ 305217/2022-4 and 151974/2024-1) and \textit{Funda\c c\~ao de Amparo \`a Pesquisa do Estado de Minas Gerais} (processes APQ-05207-23 and N. 5.16/2022).

\appendix
\section{Modified critical radius in the optical geometry}

For completeness, we analyze how the optical metric modifies the critical radius associated with circular photon orbits. Restricting again to the equatorial plane, the first integral of the null equation may be written in terms of \(u=1/r\) as
\begin{equation}
F(u)
=
-u^2+r_su^3+\frac{1+r_m^6u^6}{D^2},
\label{eq:F_u_appendix}
\end{equation}
where \(D\) is the impact parameter and \(r_m\) is the magnetic length scale introduced in Sec.~\ref{sec:math_setup}.

The critical orbit is determined by
\begin{equation}
F(u_c)=0,
\qquad
F'(u_c)=0 .
\label{eq:critical_conditions_appendix}
\end{equation}
From \(F(u_c)=0\), one obtains
\begin{equation}
D_c^2
=
\frac{1+r_m^6u_c^6}
{u_c^2(1-r_su_c)} .
\label{eq:Dc_appendix}
\end{equation}
The positivity of \(D_c^2\) requires \(u_c<1/r_s\), namely that the orbit lies outside the Schwarzschild horizon.

Differentiating Eq.~\eqref{eq:F_u_appendix} gives
\begin{equation}
F'(u)
=
-2u+3r_su^2+\frac{6r_m^6u^5}{D^2}.
\end{equation}
Substituting Eq.~\eqref{eq:Dc_appendix} into \(F'(u_c)=0\), we find
\begin{equation}
u_c
\left[
-2+3r_su_c
+
\frac{6r_m^6u_c^6(1-r_su_c)}
{1+r_m^6u_c^6}
\right]=0 .
\end{equation}
Discarding the trivial solution \(u_c=0\) and multiplying by the denominator, the criticality condition becomes
\begin{equation}
-2+3r_su_c+4r_m^6u_c^6-3r_sr_m^6u_c^7=0 .
\label{eq:critical_u_correct}
\end{equation}
Equivalently, in terms of \(r_c=1/u_c\),
\begin{equation}
2r_c^7-3r_s r_c^6-4r_m^6 r_c+3r_s r_m^6=0 .
\label{eq:critical_r_correct}
\end{equation}
Thus, after imposing both criticality conditions, the radius is determined by a seventh-degree polynomial. In general, this equation has no useful closed analytic expression for its roots. Nevertheless, the physically relevant solution is unambiguously selected by its Schwarzschild limit:
\begin{equation}
\lim_{r_m\to 0} r_c = \frac{3}{2}r_s .
\label{eq:schwarzschild_limit}
\end{equation}
This condition selects the branch continuously connected to the usual Schwarzschild photon sphere.

Introducing the dimensionless radius $x_c\equiv r_c/R$, and using \(r_s/R=2{\cal C}\), Eq.~\eqref{eq:critical_r_correct} becomes
\begin{equation}
x_c^7 - 3{\cal C}x_c^6 - 2\beta x_c + 3{\cal C}\beta=0 .
\label{eq:critical_x_correct}
\end{equation}
In the Maxwellian limit, \(\beta\to0\), the physical root is
\begin{equation}
x_c^{(0)}
=
3{\cal C},
\qquad
r_c^{(0)}
=
\frac{3r_s}{2}
=
3{\cal C}R .
\end{equation}
For small \(\beta\), the physical root can be written perturbatively as
\begin{equation}
x_c^{\rm phys}
=
3{\cal C}
\left[
1+
\frac{\beta}{(3{\cal C})^6}
+
O(\beta^2)
\right],
\end{equation}
or, equivalently,
\begin{equation}
r_c^{\rm phys}
=
\frac{3r_s}{2}
\left[
1+
\left(
\frac{r_m}{3r_s/2}
\right)^6
+
O(r_m^{12})
\right].
\label{eq:rc_perturbative}
\end{equation}
For positive \(\lambda\), the NLED correction shifts the optical critical radius outward.

The crucial question is whether the shifted critical radius lies outside the star. This requires
\begin{equation}
r_c^{\rm phys}>R,
\qquad
\text{or equivalently}
\qquad
x_c^{\rm phys}>1 .
\end{equation}
The threshold is obtained by setting \(x_c=1\) in Eq.~\eqref{eq:critical_x_correct}. This gives
\begin{equation}
\beta_{\rm ext}
=
\frac{1-3{\cal C}}{2-3{\cal C}} .
\label{eq:beta_ext}
\end{equation}
Hence, an exterior critical orbit exists only if
\begin{equation}
\beta>\beta_{\rm ext}.
\label{eq:external_condition}
\end{equation}
In model-independent form, this condition may be written as
\begin{equation}
B_s>
B_s^{\rm ext}
\equiv
\sqrt{\frac{\beta_{\rm ext}}{\lambda}} .
\label{eq:B_ext_general}
\end{equation}

For the sake of illustration, let us now specialize to the case of Euler-Heisenberg with values given in Sec.\ \ref{subsec:angle_EH}. For this model, we get
\begin{equation}
B_s^{\rm ext}
=
B_c
\left[
\frac{\beta_{\rm ext}}{2\eta_1^{\rm EH}}
\right]^{1/2}.
\label{eq:B_ext_EH}
\end{equation}
Using the representative compactness of ${\cal C}=0.2$ adopted in the phenomenological estimates, one obtains
\begin{equation}
\beta_{\rm ext}
\simeq
2.86\times10^{-1}.
\end{equation}
With \(B_c=4.41\times10^{13}\,{\rm G}\), this gives
\begin{equation}
B_s^{\rm ext}
\simeq
2.3\times10^{15}\,{\rm G}.
\label{eq:B_ext_numeric}
\end{equation}
Thus, within the corrected post-Maxwellian convention, the benchmark magnetar fields considered in the main text, \(B_s\sim10^{15}\,{\rm G}\), shift the optical critical radius outward but do not place it outside the stellar surface.
\begin{table}[ht]
\centering
\begin{tabular}{c c c c}
\hline
Class & \(B_s\,[{\rm G}]\) & \(\beta_{\rm EH}\) & \(r_c^{\rm phys}\,[{\rm km}]\) \\
\hline
ordinary pulsar      & \(10^{13}\)             & \(5.2\times10^{-6}\) & \(7.20\) \\
low-field magnetar   & \(10^{14}\)             & \(5.2\times10^{-4}\) & \(7.28\) \\
magnetar             & \(4\times10^{14}\)      & \(8.4\times10^{-3}\) & \(8.03\) \\
strong magnetar      & \(8\times10^{14}\)      & \(3.4\times10^{-2}\) & \(9.13\) \\
benchmark magnetar   & \(10^{15}\)             & \(5.2\times10^{-2}\) & \(9.59\) \\
ultra-strong magnetar & \(1.1\times10^{15}\)   & \(6.3\times10^{-2}\) & \(9.82\) \\
near-threshold field & \(2.3\times10^{15}\)    & \(2.8\times10^{-1}\) & \(11.97\) \\
exterior example     & \(2.4\times10^{15}\)    & \(3.0\times10^{-1}\) & \(12.08\) \\
\hline
\end{tabular}
\caption{Physical critical radius obtained from Eq.~\eqref{eq:critical_x_correct} for \(R=12\,{\rm km}\), \({\cal C}=0.2\), \(B_c=4.41\times10^{13}\,{\rm G}\), and \(\eta_1^{\rm EH}=5.1\times10^{-5}\). The physical root is the one continuously connected to \(r_c=3r_s/2\) as \(B_s\to0\).}
\label{tab:critical_radius_pulsars_magnetars}
\end{table}

Furthermore, by taking \(R=12\,{\rm km}\) and \({\cal C}=0.2\), it is possible to solve Eq.~\eqref{eq:critical_x_correct} numerically and select the root continuously connected to \(r_c=3r_s/2\) as \(B_s\to0\). From this, we obtain the representative values shown in Table~\ref{tab:critical_radius_pulsars_magnetars}. This comparison shows that the effect is negligible for ordinary pulsars: although the NLED correction shifts the critical radius outward, the corrected radius remains close to the Schwarzschild value and lies well inside the stellar surface. For magnetars, the shift can be larger, but with the corrected post-Maxwellian normalization the critical radius remains inside the star for fields \(B_s\sim 10^{15}\,{\rm G}\). An exterior photon-sphere-like orbit in the optical geometry appears only for stronger fields, larger than \(B_s\simeq2.3\times10^{15}\,{\rm G}\) for the compactness adopted here.

This conclusion should be interpreted with care. The same parameter \(\beta=(r_m/R)^6\) controls the size of the NLED correction at the stellar surface. The condition for the square root appearing in the emission-angle relation to remain real at the surface is $\beta<1-2{\cal C}$. For \({\cal C}=0.2\), this gives $\beta<0.6$. In the Euler-Heisenberg case, for instance, this corresponds to
\begin{equation}
B_s
\lesssim
3.4\times10^{15}\,{\rm G}.
\end{equation}
Thus, the emergence of an exterior optical critical radius occurs within the formal surface-reality range, but at fields substantially larger than the benchmark values used for the radius-inference and time-delay estimates. The robust conclusion is therefore not that all magnetars possess an exterior critical orbit, but rather that magnetar-strength fields can move the optical critical radius outward, and only the most extreme fields may push it outside the stellar surface.

In contrast, for ordinary pulsars with \(B_s\sim10^{13}\,{\rm G}\), the dimensionless parameter \(\beta\) is many orders of magnitude below \(\beta_{\rm ext}\), and the optical critical radius remains safely inside the star. Therefore, no exterior NLED critical radius is expected for ordinary pulsars in the regime considered here.


\bibliographystyle{aasjournal}
\bibliography{ref} 

\end{document}